\documentclass[twocolumn,twoside,prd]{revtex4} 

\newif\ifpdf \ifx\pdfoutput\undefined \pdffalse \else \pdftrue \fi 
\ifpdf \usepackage[pdftex]{graphicx} \else \usepackage{graphicx} \fi 
\usepackage{fancyhdr}
\usepackage{amsmath}
\usepackage{amsthm}
\newif\ifJASA \JASAfalse

\def \met {{\,/\!\!\!\!E_{T}}}

\def \figuresize {3.45in}

\def \scriptP {{\ensuremath{{\cal P}}}}
\def \twiddleScriptP {{\ensuremath{\tilde{\cal P}}}}
\def \FewKDE {{\sc FewKDE}}
\def \Sleuth {{\sc Sleuth}}
\def \Quaero {{\sc Quaero}}

\def \H {\ensuremath{{\cal H}}}
\def \SM {\ensuremath{{\text{SM}}}}
\def \D {\ensuremath{{\cal D}}}

\begin{document}


\title{Systematic Analysis of HEP Collider Data}
\author{Bruce Knuteson}
\homepage{http://mit.fnal.gov/~knuteson/}
\email{knuteson@mit.edu}
\affiliation{Massachusetts Institute of Technology}

\date{\today}

\begin{abstract}
Compelling arguments suggest the presence of new physics at energy scales that will be probed by frontier energy colliders over the next decade.  Arguments for each of the many flavors of new physics that have been proposed seem much less compelling.  The wide variety of experimental signatures by which new physics may manifest itself suggests the desirability of analyzing all high energy collider data in one systematic framework.  These proceedings describe two potentially useful pieces of such a framework:  \Sleuth\ enables a model-independent search for new high-$p_T$ physics, and \Quaero\ automates tests of particular hypotheses against high energy collider data.  A sampling of algorithmic detail is provided in the form of a procedure for choosing an optimal binning when computing likelihood ratios.
\end{abstract}

\maketitle
\tableofcontents 


\section{Context}

The audience for this talk (and these proceedings) comprises astrophysicists, cosmologists, and statisticians, in addition to high energy experimentalists.  It is therefore worth beginning by discussing the nature of high energy collider data, particularly those features that make these data amenable to the algorithms described here.  These data are collected by large, complex detectors that record on roughly a million channels the debris from the collisions of particles (protons, electrons, and their antimatter counterparts) traveling within a few hundred miles per hour of the speed of light.

\begin{figure} \includegraphics[width=\figuresize]{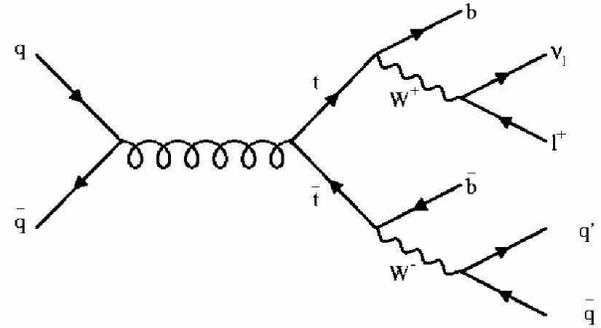} \caption{A Feynman diagram, showing the annihilation of a quark ($q$) and antiquark ($\bar{q}$), and the subsequent production and decay of a top quark ($t$) and an antitop quark ($\bar{t}$).  Time increases to the right.} \label{fig:ttbarFeynmanGraph} \end{figure}

The information contained in these million channels of electronics is reduced through a series of steps to roughly one dozen numbers, corresponding to the energies and directions (polar and azimuthal angles) of the elementary objects emerging from the collision.  This severe reduction in detail facilitates a direct connection to the underlying theory.  The underlying theory is most easily understood graphically in terms of Feynman diagrams, an example of which is shown in Fig.~\ref{fig:ttbarFeynmanGraph}.  Our detectors and algorithms (imperfectly) reconstruct the outgoing particles in collisions like that depicted in Fig.~\ref{fig:ttbarFeynmanGraph}.  The goal is to figure out, from the debris of trillions of particle collisions, the rules corresponding to graphs such as that shown in Fig.~\ref{fig:ttbarFeynmanGraph}:  rules for what types of graphs can be drawn, and rules for calculating observable quantities from them.  In doing so, we infer from measurements on scales of meters the laws of Nature on scales of $10^{-16}$ meters and below.

The theoretical context in which we work is grounded in the standard model of particle physics, which predicts the results of nearly all experiments performed to date with extraordinary accuracy --- and in many cases also with extraordinary precision.  This standard model represents a canonical reference model, the null hypothesis in our field. 

\begin{figure} \includegraphics[width=\figuresize]{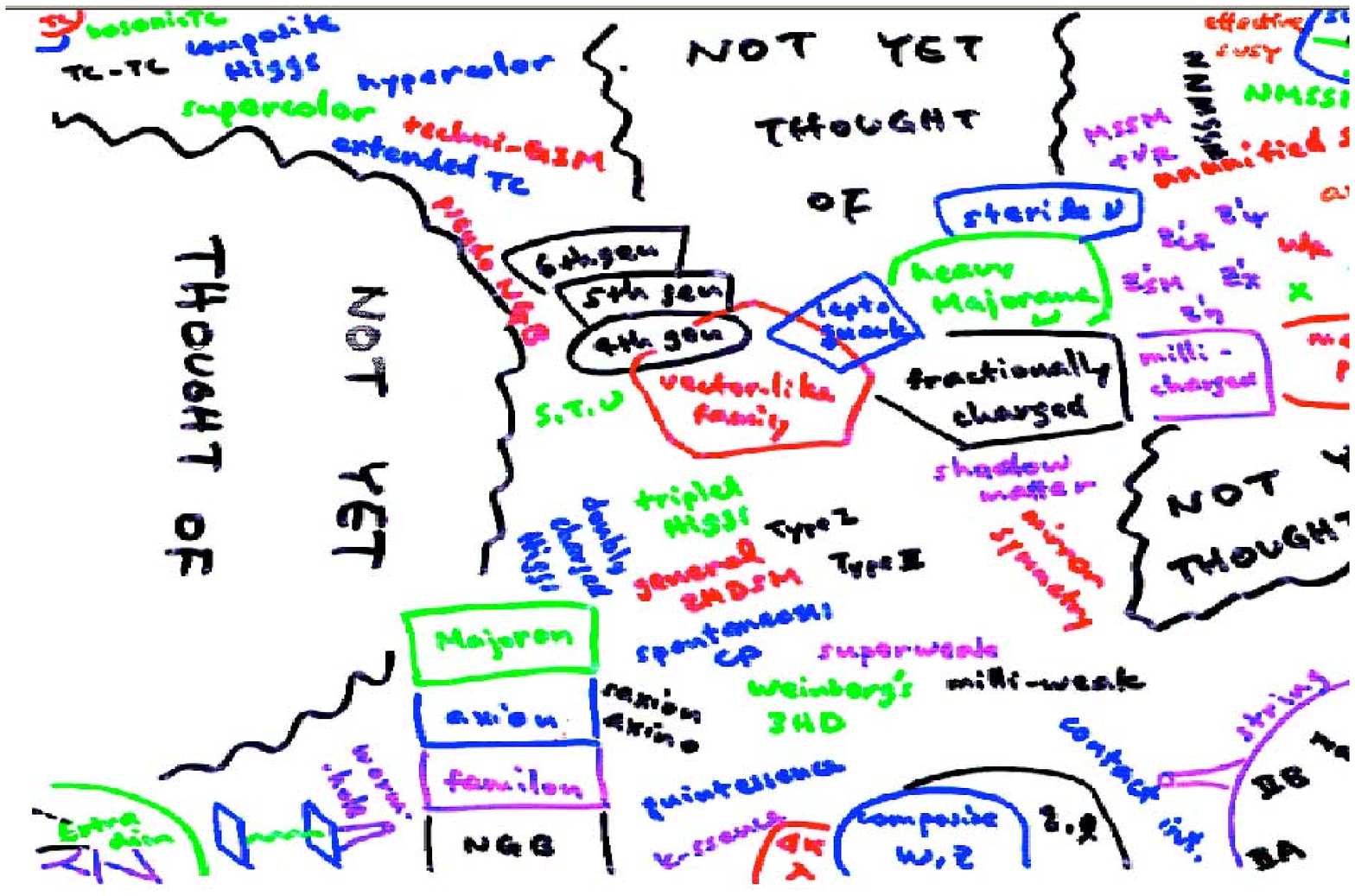} \caption{The theoretical landscape, as depicted in the summary talk of this year's Lepton Photon conference~\cite{TheoryScape}.} \label{fig:TheoreticalLandscape} \end{figure}

The theoretical landscape beyond the standard model is much less clear.  Hundreds of different scenarios have been proposed, each containing many parameters.  The lack of clarity in this picture is nicely captured in a slide shown during the summary talk of Lepton Photon 2003, reproduced in Fig.~\ref{fig:TheoreticalLandscape}.

\section{\Sleuth}

The jumbled theoretical landscape in Fig.~\ref{fig:TheoreticalLandscape}, reflecting the plethora of possible extensions to the standard model, calls into question the paradigm currently being used to explore that landscape.  At present roughly one graduate student is consumed for each model tested.  

An alternative way to proceed is to systematically search for any evidence of new physics that lies in the data, in a manner that is as model-independent as possible.  A prescription for doing this is an algorithm called \Sleuth, used by the D\O\ experiment in Run~I to search a large subset of their data~\cite{SleuthPRD1:Abbott:2000fb, SleuthPRD2:Abbott:2000gx, SleuthPRL:Abbott:2001ke, KnutesonThesis}.  

One of many problems faced when searching for new physics in such a directionless landscape is how to take into account the large space of possible signatures that could appear when computing a final measure of the significance of any particular result.  If many students look at many plots over an extended period of time, fluctuations at the level of three or more standard deviations are bound to appear simply from the fact that thousands of bins in various histograms have been considered.  The difficulty in computing this {\em trials factor}, the number of possible places that an interesting signal could have appeared, has hamstrung several previous search efforts that have attempted to base themselves on signatures rather than models.  A rigorous accounting of the trials factor is crucial to any model-independent search; \Sleuth\ is one of the few algorithms currently on the market that is able to compute this trials factor rigorously and explicitly.  The H1 Collaboration has developed an algorithm in similar spirit for HERA physics~\cite{SleuthH1}.

Key to a rigorous computation of the trials factor is defining --- before the data is collected --- the interestingness of any particular signature that might be seen in those data.  \Sleuth\ is able to do this by making three well-justified assumptions.  
\begin{enumerate}
\item The data can be categorized into exclusive final states in such a way that any signature of new physics is apt to appear predominantly in one of these final states.  
\item New physics will appear with objects at high transverse momentum ($p_T$) relative to standard model and instrumental background.  
\item New physics will appear as an excess of data over background.  
\end{enumerate}
The \Sleuth\ algorithm consists of three steps, following these three assumptions.

In the first step, all of the collisions are partitioned into exclusive final states.  The objects used to categorize these final states are high-$p_T$ and isolated electrons ($e$), muons ($\mu$), taus ($\tau$), photons ($\gamma$), jets ($j$), b-tagged jets ($b$), and missing transverse energy ($\met$).  

The second step of the algorithm defines a low-dimensional variable space for each final state.  In the Run~I implementation of \Sleuth, the variables used were 
\begin{itemize}
\item the summed transverse momentum of any leptons in the event ($\sum{{p_T}^{e/\mu/\tau}}$);
\item the missing transverse energy ($\met$), if significant in the event;
\item the summed transverse momentum of any electroweak gauge bosons in the event ($\sum{{p_T}^{W/Z/\gamma}}$); and 
\item the summed transverse momentum of any jets in the event ($\sum{{p_T}^{j}}$).
\end{itemize}
The Run II algorithm is simplified enormously by considering only a single variable,
\begin{itemize}
\item the summed transverse momentum of all objects in the event ($\sum{p_T}$).
\end{itemize}
New high-$p_T$ physics is best searched for by systematically looking for new physics at high $p_T$.  

 The algorithm's third step involves searching for regions in which more events are seen in the data than expected from standard model and instrumental background.  This search is performed in the variable space defined in the second step of the algorithm, for each of the exclusive final states defined in the first step.  

The details of the search are somewhat involved, but both the input and output are exceptionally simple.  For each final state, the input is simply the events seen in the data, and the expected background.  The steps of the search can be sketched as follows.
\begin{figure} \includegraphics[width=\figuresize]{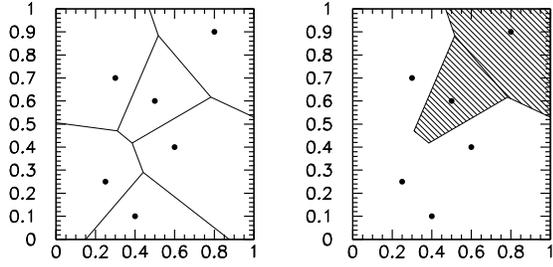} \caption{A Voronoi diagram with seven data points (black dots) in a unit square (left).  The Run~I \Sleuth\ algorithm considers regions that are unions of these cells, such as the shaded region (right).} \label{fig:VoronoiDiagram} \end{figure}

\begin{itemize}
\item The variable space is transformed into the unit box --- the unit interval in one dimension, unit square in two dimensions, unit cube in three dimensions, and unit hypercube in four dimensions.  
\item The notion of regions about sets of data points is rigorously defined using the concept of Voronoi diagrams, borrowed from the field of computational geometry.  Figure~\ref{fig:VoronoiDiagram} shows an example of a unit square containing seven data points, shown as black dots.  The perpendicular bisectors of line segments connecting each pair of data points connect to form the Voronoi diagram.
\item The interestingness of any particular region (or union of such regions) in Fig.~\ref{fig:VoronoiDiagram} is the Poisson probability that the background in that region would fluctuate up to or above the observed number of events in that region.  
\item The most interesting region ${\cal R}$ is found using a search heuristic to explore the space of potentially interesting regions.
\item Pseudo experiments are performed to determine the fraction ${\cal P}$ of hypothetical similar experiments in which something more interesting than ${\cal R}$ would be seen.  Here the fact that many different places have been considered is rigorously and explicitly accounted for.  \Sleuth\ and its H1 analogue appear to be the only algorithms currently on the market for frontier energy collider physics that compute this trials factor completely and systematically.
\item The results from all final states considered are then combined to form $\twiddleScriptP$, which quantifies the interestingness of the most interesting region observed in the data, accounting for the fact that many final states have been considered.  
\end{itemize}

The Run II algorithm is trivial by comparison.  In the single variable $\sum{p_T}$, semi-infinite regions are defined with a lower bound at each data point.  The definition of interestingness, running of pseudo experiments, and calculation of $\scriptP$ and $\twiddleScriptP$ proceed as above.

The output of the algorithm is the most interesting region ${\cal R}$ observed in the data, and a number $\twiddleScriptP$ that quantifies the interestingness of ${\cal R}$.  $\twiddleScriptP$ is a number between zero and unity, pulled from a uniform distribution on the unit interval if the data comes from background alone, and expected to be small if the data contain a hint of new physics.  A reasonable threshold for discovery is $\twiddleScriptP \lesssim 0.001$, which corresponds loosely to the {\em de facto} $5\sigma$ standard in our field after the trials factor is accounted for.~\footnote{The threshold of ${\twiddleScriptP} \lesssim 10^{-3}$ follows directly from our field's standard discovery threshold of five standard deviations in the following manner.  We recall that $5 \sigma$ corresponds to a probability of roughly $10^{-7}$.  In a collaboration the size of CDF there are $\approx 100$ graduate students.  Each student works 2 years on his analysis, and makes 1 interesting plot per week.  100 students $\times$ 50 weeks/year $\times$ 2 years = $10^4$ different ``things'' that are looked at.  $10^{-7} \times 10^4 = 10^{-3}$, which corresponds to roughly 3 standard deviations.  The desire to see a $5\sigma$ effect is therefore a desire to see a $3\sigma$ effect when the accounting is done for all possible places a signal could have appeared, but did not.  \Sleuth\ includes this accounting (performed much more rigorously than in this footnote) in the calculation of $\twiddleScriptP$.}

Two questions must now be asked:
\begin{itemize}
\item Will \Sleuth\ find nothing if there is nothing to be found?
\item Will \Sleuth\ find something if there is something to be found?
\end{itemize}
The answer to the first is ``yes,'' by construction~\footnote{Spurious signals will of course be seen if \Sleuth\ is provided improperly modeled backgrounds.  \Sleuth\ directly addresses the issue of whether an observed hint is due to a statistical fluctuation; it is unable to address systematic mismeasurement or incorrect modeling (but quite useful in bringing these to your attention).}.  The answer to the second depends to what extent the new physics waiting to be uncovered satisfies the three assumptions on which \Sleuth\ is based.  

\begin{table}[htb]
\centering
\begin{tabular}{c|c|c|c}
& \multicolumn{3}{c}{Omitted} \\
         & $WW$, $t\bar{t}$ & $t\bar{t}$ & none \\ \hline
Final state  	& \multicolumn{3}{c}{$\scriptP$} \\ \hline
$e\mu\met$   	& {$\mathbf{2.4\sigma}$} & $1.1\sigma$ & $1.1\sigma$\\
$e\mu\met j$  	& $0.4\sigma$	& $0.1\sigma$ & $0.1\sigma$ \\
$e\mu\met jj$ 	& {$\mathbf{2.3\sigma}$} & {$\mathbf{1.9\sigma}$} & $0.5\sigma$ \\
$e\mu\met jjj$  & $0.3\sigma$	& $0.2\sigma$ & $-0.5\sigma$ \\ \hline
\raisebox{-.6ex}{$\twiddleScriptP$} & $1.9\sigma$ & $1.2\sigma$ & $-0.6\sigma$ \\ 
\end{tabular}
\caption{Summary of a \Sleuth\ sensitivity study on the $e\mu\met$, $e\mu\met j$, $e\mu\met jj$, and $e\mu\met jjj$ final states.  When the standard model processes $WW$ and $t\bar{t}$ are omitted from the background estimate (second column), \Sleuth\ identifies a region of excess in the $e\mu\met$ and $e\mu\met jj$ final states (with $\scriptP=2.4\sigma$ and $2.3\sigma$, respectively), presumably indicating the true presence of $WW$ and $t\bar{t}$ in the data.  When the standard model process $WW$ is included and $t\bar{t}$ omitted (third column), \Sleuth\ identifies a region of excess in the $e\mu\met jj$ final state (with $\scriptP=1.9\sigma$), presumably indicating the true presence of $t\bar{t}$ in the data.  With all standard model processes included to search for new physics (third column), \Sleuth\ indicates that 72\% ($\twiddleScriptP=-0.6\sigma$) of background-only hypothetical similar experiments would have produced a region more interesting than the most interesting region observed in these data.}
\label{tbl:SleuthStudy}
\end{table}

Although no general answer can be given to this second question, an answer can be given for any specific case.  Such a specific case is summarized in Table~\ref{tbl:SleuthStudy}~\cite{SleuthPRD1:Abbott:2000fb}.  Events containing an energetic electron, muon, and possibly other objects ($e\mu X$) are considered.  In a first pass, standard model $WW$ and $t\bar{t}$ production are omitted from the background estimate to see if \Sleuth\ is able to find evidence of these processes in D\O\ Run~I data, and the result ${\cal P}$ obtained in each final state (translated into units of standard deviations) is shown.  \Sleuth\ finds ${\cal P}= 2.4\sigma$ and $2.3\sigma$ in the final states $e\mu\met$ and $e\mu\met jj$; these excesses correspond (presumably) to the true presence of $WW$ and $t\bar{t}$ in these data.  For comparison, a dedicated search for $WW$ in Run~I at CDF~\cite{Abe:1997dw} resulted in 5 events observed on a background of $1.2\pm0.3$, corresponding to a significance of $2.3\sigma$; and a dedicated search in $e\mu X$ for $t\bar{t}$ by D\O\ in Run~I~\cite{Abachi:1997re} resulted in 5 events observed on a background of $1.4\pm0.4$, corresponding to significance of $2.1\sigma$.  

The quantity $\scriptP$ obtained from \Sleuth\ really should not be directly compared to the result of a dedicated search, since the two techniques are intended for very different problems:  dedicated searches are clearly preferred if there are well-defined, compelling things to be found, while \Sleuth\ provides an alternative strategy in their absence.  This example nonetheless provides useful intuition for \Sleuth's performance on a difficult test.

In a second pass, standard model $WW$ production is included in the background estimate, with standard model $t\bar{t}$ production still omitted, to see whether \Sleuth\ could find evidence of $t\bar{t}$ in these data.  The results obtained are shown in the third column of Table~\ref{tbl:SleuthStudy}, with the excess in the $e\mu\met jj$ final state corresponding (presumably) to the actual presence of $t\bar{t}$ in these data.  The slight indications of excess in these examples clearly fall well short of that needed to make a discovery claim; as indicated above, these are difficult tests.  

With all backgrounds included and \Sleuth\ used to search for new physics in the fourth column of Table~\ref{tbl:SleuthStudy}, a null result is obtained.  The use of \Sleuth\ to analyze roughly thirty additional final states at D\O\ in Tevatron Run~I resulted in no evidence of new physics~\cite{SleuthPRD1:Abbott:2000fb, SleuthPRD2:Abbott:2000gx, SleuthPRL:Abbott:2001ke, KnutesonThesis}.

A general model-independent search in similar spirit~\cite{SleuthH1} has recently been presented by the H1 collaboration at the 2003 European Physical Society meeting in Aachen, Germany.  It will be interesting to continue to watch their $\mu j \nu$ final state in HERA Run II.

\section{\Quaero}

The first hint of new physics at Tevatron Run II may come from a model-independent search.  Once such a hint is found, it must be interpreted in terms of an underlying physical theory.  This interpretation would clearly be facilitated by some means of quickly and efficiently testing the predictions of many different hypotheses against the data.  \Quaero\ (Latin for ``I search for,'' or ``I seek'') is an algorithm designed for this purpose.

Present practice for testing hypotheses against collider data can be improved upon in several respects.  A personal wish list for conducting analyses on high energy collider data includes:
\begin{itemize}
\item Reducing the time spent to perform an analysis from two years of one graduate student's life to roughly an hour of CPU time.  Achieving this would represent a reduction in the time it takes to perform an analysis by a factor of $10^4$.
\item Reducing human bias that invariably creeps into analyses on complex data sets.
\item Allowing the publication of data in their full dimensionality, unrestricted by the two dimensions of a sheet of paper.
\item Providing an alternative to exclusion contours.  The exclusion plots often shown make it difficult to understand exactly what model is being tested, together with all assumptions that are made, and difficult to tell what the data have to say about a model that does not lie in that two-dimensional space.  
\item Automating the optimization of analyses, to ensure the data are used to their fullest.
\item Rigorously propagating systematic errors in an intuitive, straightforward, and rigorous way.  
\item Combining results among correlated experiments in a manner that requires as few {\em ad hoc} prescriptions as possible.
\item Increasing the robustness of our scientific results by using a high-level analysis algorithm that has been validated on hundreds of previous analyses.
\item All of this on the web.
\end{itemize}

A first pass of such an algorithm has been achieved.  With the posting of an article entitled ``Search for New Physics Using \Quaero:  A General Interface to D\O\ Event Data"~\cite{QuaeroPRL:Abazov:2001ny}, D\O\ has made a subset of data collected in Tevatron Run~I available on the web at \url{http://quaero.fnal.gov/} since June 2001.

\begin{figure} \includegraphics[width=\figuresize]{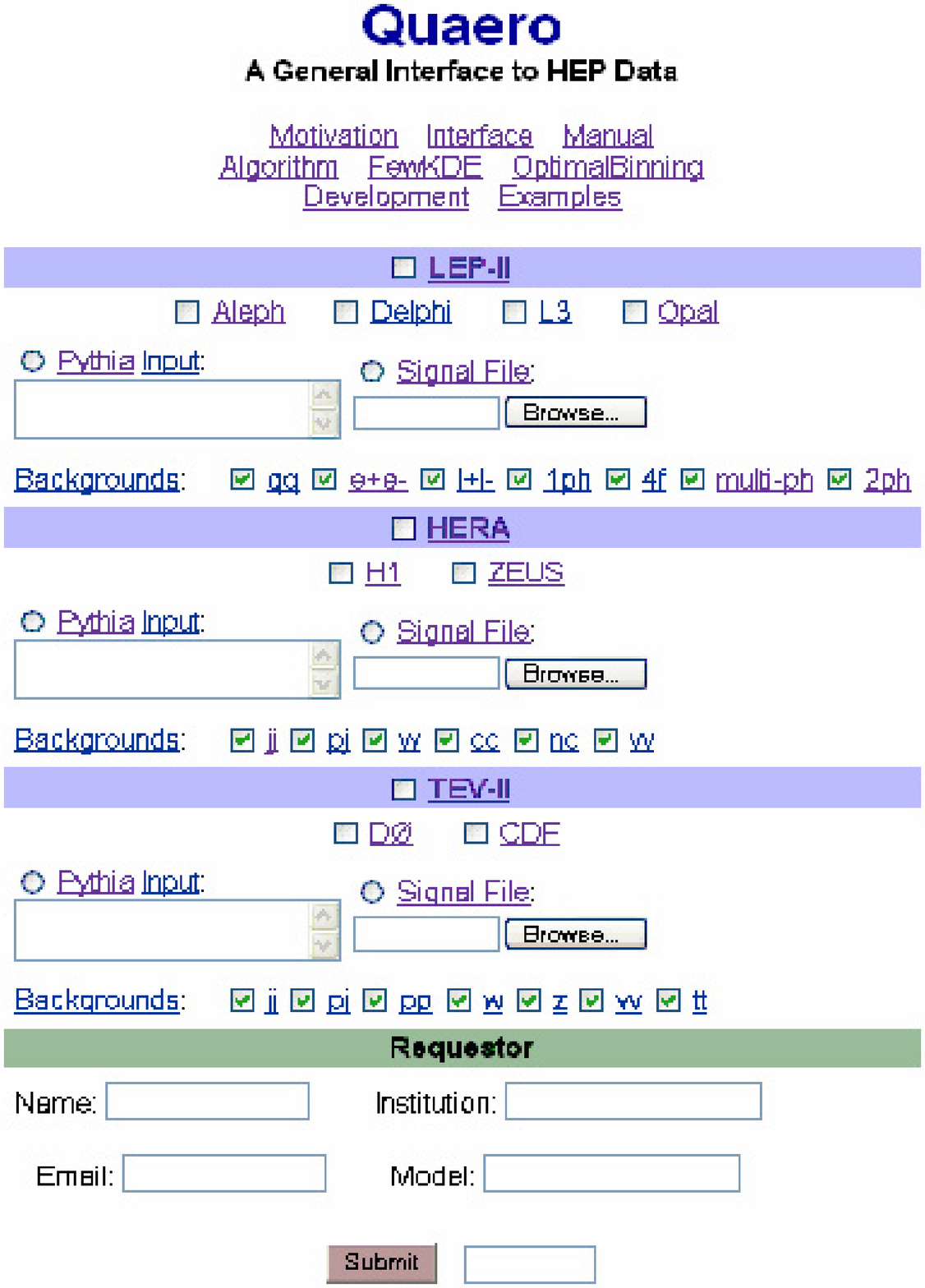} \caption{The \Quaero\ web page under development for frontier energy collider data.} \label{fig:QuaeroWebPage} \end{figure}

Astrophysicists have become accustomed to polished interfaces to their data; the web page served up by the Sloan Digital Sky Survey at \url{http://www.sdss.org/} is one of many examples.  Those in the audience with this image in mind are bound to be disappointed by the look and feel of Fig.~\ref{fig:QuaeroWebPage} --- high energy physics is at least a decade behind the astrophysics and astronomy communities on this front.

The essential \Quaero\ interface, devoid of adornment, is displayed in Fig.~\ref{fig:QuaeroWebPage}.  A physicist with a particular hypothesis ${\cal H}$ to test against high energy collider data should be able to provide his hypothesis in the form of the events his model predicts --- either as input to an event generator, or as a file with the events themselves.  These events (the ``signal''), together with whatever standard model processes (``backgrounds'') he wishes to include, define his hypothesis for how Nature works at very small distance scales.

After providing his name and the email address to which the result should be sent, the physicist clicks ``Submit.''  \Quaero\ then performs the complete analysis, taking into full account the expert knowledge gleaned within the collaboration and packaged into code, and returns a single number, quantifying the extent to which the data (dis)favor the hypothesis relative to the standard model.  \Quaero\ also provides a number of plots showing in detail how the analysis was performed.  Far from being a black box, \Quaero\ arguably provides a much more transparent view into how analyses are performed than our standard publications.

The \Quaero\ algorithm itself is relatively simple, involving a few straightforward steps.
\begin{itemize}
\item{The events predicted by the hypothesis $\H$ are run through the detector simulation appropriate for each experiment.}
\item{Events from $\H$, the standard model ($\SM$), and the data ($\D$) are partitioned into exclusive final states.  Speaking loosely, these final states are orthogonal (no event belongs to more than one final state) and complete (every event belongs to a final state).}
\item{Variables are chosen automatically within each final state.}
\item{A binning is chosen automatically within the variable space in each final state.}
\item{A binned likelihood is calculated within each final state.}
\item{Results from different final states are combined.}
\item{Results from different experiments are combined.}
\item{Systematic errors are integrated numerically.}
\item{The result returned is a likelihood ratio, 
\begin{equation}
{\cal L}(\H) = \frac{p(\D|\H)}{p(\D|\SM)}.
\end{equation}
}
\end{itemize}

In order to provide a feeling for the details of the algorithm within the space constraints of these proceedings, one piece of the algorithm is highlighted: automatic choice of binning.

\section{Optimal Binning for Likelihood Ratios}

A binned likelihood provides a robust yet sensitive method for discriminating between two hypotheses.  But how should the bins be chosen?  Somewhat surprisingly, the literature does not yet appear to contain a satisfactory general prescription for choosing an optimal binning.  This section suggests such a prescription, investigates its implications in several limiting cases, and provides examples of its use.

\begin{figure} \includegraphics[width=\figuresize]{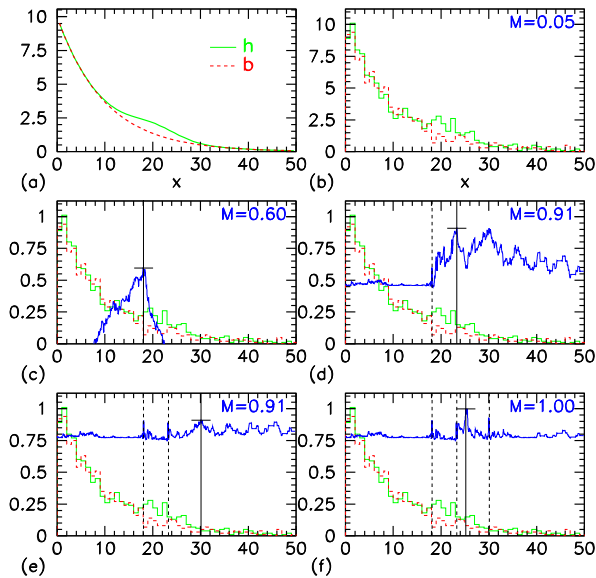} \caption{Placement of bins in a toy example: a bump on a falling exponential.  The true (unknown, analytic) distributions $h(x)$ (solid, green) and $b(x)$ (dashed, red) are shown in (a); our knowledge of these distributions, in the form of 1000 Monte Carlo points drawn from each, is shown in (b).  In this case $b(x)$ is a simple exponential, with $h(x)$ adding a Gaussian bump centered at $x=20$.  The vertical axes in (a) and (b) represent the number of events expected per unit of $x$.  Sequential placement of bin edges is shown in (c)--(f), with the figure of merit ${\cal M}$ on the vertical axis.} \label{fig:Example1} \end{figure}

Figures~\ref{fig:Example1}(a) and (b) show a typical problem.  Predicted (analytic) distributions from two hypotheses $h$ and $b$ are shown in Fig.~\ref{fig:Example1}(a).  Units on the vertical axis are the number of predicted events per unit of $x$, the observable shown on the horizontal axis.  Often the analytic form of the predictions are not known, however; knowledge of the predictions from $h$ and $b$ come in the form of an ensemble of Monte Carlo events, whose statistics are limited by the complexity of the simulation required for each event.

It is desired to perform an experiment to collect data $d$ to determine whether hypothesis $h$ or $b$ is the more accurate description of Nature.  The number we would like to determine is the likelihood ratio $p(d|h)/p(d|b)$ --- in words, the probability of obtaining the data $d$ assuming the correctness of the hypothesis $h$, divided by the probability of obtaining the data $d$ assuming the correctness of the hypothesis $b$.  

Given the predictions from $h$ and $b$ shown in Fig.~\ref{fig:Example1}(b), how should this likelihood ratio be computed?  If the predictions $h(x)$ and $b(x)$ were known as analytic functions of $x$, as in Fig.~\ref{fig:Example1}(a), an unbinned likelihood could be calculated.  But the analytic forms $h(x)$ and $b(x)$ are not known.  Constructing smooth distributions $h(x)$ and $b(x)$ from Monte Carlo points using smoothing techniques is possible, but the final answer is often unfortunately sensitive to the details of how this smoothing is performed.   The only reasonable option appears to involve the introduction of bins, and the computation of a binned likelihood.  

But then how should the bins be set?  The bins must clearly be fine enough to probe the difference in shape between the two distributions; the bins must just as clearly be large enough that an accurate prediction is obtained for the number of events $h_k$ and $b_k$ in each bin $k$.  The issue at hand is not only how many bins to use, but also where to place their edges.

There is no unique solution to this problem.  The best one can do is to define a reasonable prescription for choosing an optimal binning --- in effect, by suggesting some reasonable definition of ``optimal'' --- and demonstrate its reasonable behavior on a variety of examples.

The prescription suggested here involves defining a figure of merit ${\cal M}$ by
\begin{multline}
\label{eqn:FigureOfMeritComplicated}
{\cal M} =  \sum_{d_1=0}^\infty \sum_{d_2=0}^\infty \cdots 
 \left( \prod_k{ p(d_k|p(h_k))  } \right) \times \\
 \log\left( \prod_k \frac{p(d_k|p(h_k))}{p(d_k|p(b_k))}\right)  \\
 + \left( h \leftrightarrow b \right) - {\cal P}.
\end{multline}
In words, ${\cal M}$ is the evidence the experiment is expected to provide in favor of $h$ if $h$ is correct, plus the evidence the experiment is expected to provide in favor of $b$ is $b$ is correct.  The definition of ``evidence'' here, adopted from Ref.~\cite{Jaynes}, is the logarithm of the likelihood ratio; ``expected'' is defined in terms of an average over an ensemble of hypothetical experiments, where the correctness of either $h$ or $b$ is assumed in weighting the possible outcomes.  

The initial sum in Eq.~\ref{eqn:FigureOfMeritComplicated} is over all possible outcomes of the experiment:  the number of data events $d_k$ in each bin $k$ is allowed to vary between zero and infinity.  The factor $\prod_k{ p(d_k|p(h_k))}$ weights each outcome by the probability of its occurrence, assuming the correctness of $h$.  Here $p(h_k)$ is our knowledge of the number of events predicted by $h$ in bin $k$; we might have $p(h_k)$ in the form of a Gaussian with mean 7 and width 1.2 if the number of events predicted by $h$ in bin $k$ were $7\pm1.2$.  The factor $\log\left( \prod_k \frac{p(d_k|p(h_k))}{p(d_k|p(b_k))}\right)$ is the evidence obtained in favor of $h$ in this outcome.  To this is added a similar term with $h$ and $b$ swapped; the second term $(h\leftrightarrow b)$ is the expected evidence in favor of $b$ if $b$ is correct.  The third term ${\cal P}$ is a penalty term, which provides the stopping condition for the algorithm's placement of bins.  

\begin{figure} \includegraphics[width=\figuresize]{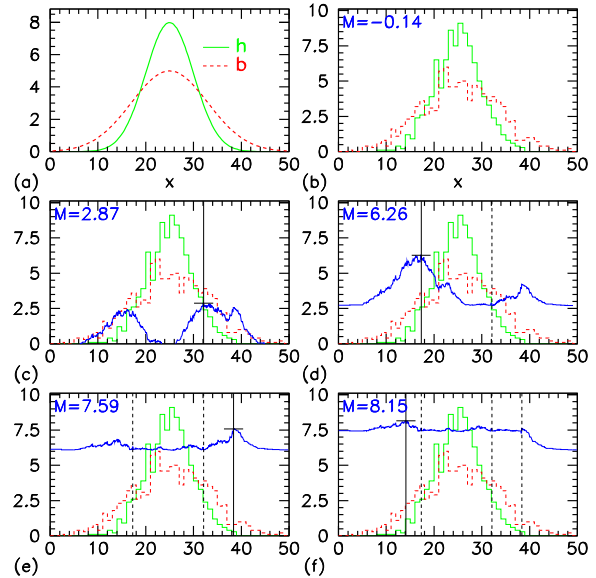} \caption{Placement of bins in a toy example: two Gaussians of different widths.  The true (unknown, analytic) distributions $h(x)$ (solid, green) and $b(x)$ (dashed, red) are shown in (a); our knowledge of these distributions, in the form of 1000 Monte Carlo points drawn from each, is shown in (b).  In this case $b(x)$ is a Gaussian centered at 25 with width 8; $h(x)$ is a Gaussian with the same mean and width 5.  The vertical axes in (a) and (b) represent the number of events expected per unit of $x$.  Sequential placement of bin edges is shown in (c)--(f), with the figure of merit ${\cal M}$ on the vertical axis.} \label{fig:Example2} \end{figure}

Figures~\ref{fig:Example1} and~\ref{fig:Example2} show how this figure of merit ${\cal M}$ can be used to determine the placement of bins.  Figure~\ref{fig:Example1}(a) shows the true (analytic and unknown) predictions $h(x)$ and $b(x)$ from the hypotheses $h$ and $b$ in the observable $x$.  Figure~\ref{fig:Example1}(b) shows our knowledge of the predictions of these two hypotheses, in the form of one thousand Monte Carlo points drawn from the true (unknown) distributions $h(x)$ and $b(x)$.  Using a single bin from 0 to 50 in $x$, the figure of merit computed using Eq.~\ref{eqn:FigureOfMeritComplicated} for the points shown in Fig.~\ref{fig:Example1}(b) is ${\cal M} = 0.05$.  

Figures~\ref{fig:Example1}(c)--(f) show the successive placement of bin edges.  In these plots the vertical axis has units of expected evidence; the predictions of $h$ and $b$ are superimposed with arbitrary scale.  At each value of $x$, the dark (jagged, blue) curve shows the figure of merit ${\cal M}$ if a bin edge is placed at that point.  In Fig.~\ref{fig:Example1}(c), the maximum of this curve is obtained with a bin edge placed at $x=18$; this raises the figure of merit to ${\cal M} = 0.60$.  Placing a bin edge at this point results in one bin stretching from 0 to 18, and one bin reaching from 18 to 50.  Figure~\ref{fig:Example1}(d) shows this process repeated, the figure of merit calculated for each possible location of a second bin edge.  Maximizing the expected evidence in the dark (jagged, blue) curve requires placement of a bin edge at $x=23$.  This placement leaves three bins: [0--18], [18--23], and [23--50].  Figures~\ref{fig:Example1}(e) and (f) show the placement of two more bin edges, at $x=30$ and at $x=25$.  Further placement of bin edges decreases the figure of merit ${\cal M}$, so the algorithm halts.

A second example is shown in Fig.~\ref{fig:Example2}.  Figure~\ref{fig:Example2}(a) shows the true (analytic and unknown) predictions $h(x)$ and $b(x)$, both Gaussians with identical mean and area but different widths.  One thousand Monte Carlo points pulled from each of $h(x)$ and $b(x)$ are shown in Fig.~\ref{fig:Example2}(b).  The use of a single bin from 0 to 50 results in a figure of merit of ${\cal M} = -0.14$; a negative value is obtained because the total number of events predicted by $h$ and $b$ in this single encompassing bin is the same (the Gaussians have equal area), and the penalty term ${\cal P}$ in Eq.~\ref{eqn:FigureOfMeritComplicated} drives the figure of merit ${\cal M}$ negative.  In Fig.~\ref{fig:Example2}(c)--(f), the units of the vertical axes are expected evidence, with the predictions of $h$ and $b$ again superimposed.  Notice the difference in vertical scale between Figs.~\ref{fig:Example1}(c)--(f) and Figs.~\ref{fig:Example2}(c)--(f); the evidence we expect the experiment to provide in favor of $h$ relative to $b$ (or vice versa) is clearly much larger in the example of Fig.~\ref{fig:Example2}.

A first bin edge is placed in Fig.~\ref{fig:Example2}(c), the figure of merit ${\cal M}$ computed as the bin edge's position is scanned in $x$, resulting in the dark (jagged, blue) curve.  As expected by looking at the true distributions for $h$ and $b$, the algorithm prefers bin placement at $x\approx 15$ or $x\approx 35$, where the analytic predictions for $h$ and $b$ cross in Fig.~\ref{fig:Example2}(a).  In Fig.~\ref{fig:Example2}(c), placement of a bin edge at $x=32$ is slightly favored.  The first bin edge is placed at this point, resulting in one bin ranging from 0 to 32, and a second bin covering 32 to 50.  The process is repeated, with the expected evidence curve shown in Fig.~\ref{fig:Example2}(d), and a second bin is placed at $x=17$, raising the total figure of merit to ${\cal M} = 6.26$.  Figures~\ref{fig:Example2}(e) and (f) show the process repeated twice more, raising the total figure of merit to ${\cal M} = 8.15$.  The algorithm places eight additional bin edges in the regions $x\approx 20$ and $x \approx 30$ before halting.  

The algorithm's performance in these two cases is remarkably intuitive.  In the first example, the procedure nicely carves out the region around the bump that $h(x)$ shows relative to $b(x)$ in Fig.~\ref{fig:Example1}(a), correctly ignoring the bulk of the distribution at $x<10$ and the tail at $x>30$.  In the second example, the algorithm systematically works from side to side in Fig.~\ref{fig:Example2}, from the right of the mean to the left of the mean and back, doggedly separating regions in which $h$ predicts more events than $b$ from regions in which $b$ predicts more than $h$.

The algorithm presented here has at least two multivariate generalizations.  One option iteratively places bin edges in the form of hyperplanes parallel to the variable axes, creating a grid in the multidimensional space.  In some cases this may be an acceptable approach, but the resulting rectangular bins are too constrained in shape to adequately handle an arbitrary multidimensional problem.  An (improved) alternative is to use kernel density estimation to first reduce the problem to a single dimension, enabling the application of the one-dimensional binning algorithm just described.

\section{\FewKDE}

Standard kernel estimation involves placing bumps of probability, typically in the form of Gaussian kernels, around each Monte Carlo point.  Summation of kernels placed around each of an ensemble of Monte Carlo points forms the density estimate.  

In this standard approach, the evaluation of the density at any particular point requires the evaluation of a Gaussian centered at each of of the $N_{MC}$ Monte Carlo points.  The time cost of evaluating this density estimate at each of the points used to generate the estimate thus grows as ${\cal O}(N_{MC}^2)$, which becomes prohibitive when dealing with samples of $\gtrsim 10^4$ Monte Carlo points.  Application to high statistics Tevatron and future LHC analyses is facilitated by noting that distributions derived from four-vector quantities of final state objects in high-$p_T$ collider physics can be approximated satisfactorily by the sum of just a few Gaussians.

An algorithm called \FewKDE\ has been introduced with the generally featureless nature of our distributions in mind, where ``\FewKDE'' is shorthand for ``kernel density estimation with few kernels.''  The parameters of the few Gaussians are chosen to provide the best fit to the data.  A novel technique is employed to appropriately handle the types of hard physical boundaries (such as $p_T>0$) that exist in commonly considered distributions.  

\section{Summary}

These proceedings have briefly sketched a method allowing the systematic analysis of high energy collider data.  After briefly providing the experimental and theoretical contexts of frontier energy collider data to the statisticians, astrophysicists, and cosmologists in the audience, a direct solution to a few of the problems we face in the analysis of those data has been described.  

Given the variety of possible forms physics beyond the standard model may take, the question of how to search for something when we know only vaguely what it is we are searching for becomes acute.  \Sleuth\ is an algorithm that accomplishes this in a rigorous and systematic way, enabling a model-independent search for new high-$p_T$ physics.

Once a hint of new physics is observed, data understood in the context of a systematic search must be interpreted in terms of the underlying physical theory.  Accomplishing this requires a procedure for quickly and efficiently testing particular hypotheses against the data.  \Quaero\ provides a qualitatively new medium for facilitating this interpretation.  

In order to provide a feeling for one of several algorithmic pieces introduced in the development of \Quaero, a procedure for optimally choosing a binning for the computation of a binned likelihood ratio has been described.  Generalization to the multivariate case makes use of \FewKDE, a time-saving variant of the standard procedure for kernel density estimation.

It is our hope that the ideas presented here, developed for a particular problem within high energy physics, may lend themselves to many other problems in the physical sciences.

\section{Acknowledgments}

The {\sc phystat 2003} conference organizers and SLAC provided a wonderful opportunity to share techniques and ideas in a creative atmosphere.  Mark Strovink, Hugh Montgomery, Greg Landsberg, Dave Toback, and others on the D\O\ collaboration contributed to the development of \Sleuth\ and \Quaero, and their application to high energy collider data.  The application of \Sleuth\ and \Quaero\ to data has been further advanced by Daniel Whiteson at D\O, Marcello Maggi at ALEPH, Andr\'e Holzner at L3, Sascha Caron at H1, and others.  Additional invaluable logistical assistance has been provided by many, including Bolek Pietrzyk, Jacques Boucrot, and Mark Oreglia.  Khaldoun Makhoul and Jang Woo Lee (MIT) have aided the implementation of the new \Quaero\ and \Sleuth\ algorithms.  Hannu Miettinen, Paul Padley, David Scott, and Sang-Joon Lee (Rice University) have assisted in the testing of \FewKDE\ and other issues related to kernel density estimation.  Henry Frisch and Mel Shochet (University of Chicago) have provided helpful discussions.

This work has been funded by a Department of Defense National Defense Science and Engineering Graduate Fellowship at the University of California, Berkeley; an International Research Fellowship from the National Science Foundation (INT-0107322); a Fermi/McCormick Fellowship at the University of Chicago; and Department of Energy grant DE-FC02-94ER40818.

\bibliography{phystat2003}

\end{document}